\documentclass[twocolumn,superscriptaddress]{revtex4}
\usepackage{amsmath,graphicx,epsfig}

\begin{document}

\title{Observation of scalar nuclear spin-spin coupling in van der Waals molecules} 

\author{Micah P.\ Ledbetter}\email{ledbetter@berkeley.edu}
\affiliation{Department of Physics, Princeton University, Princeton,
New Jersey 08544, USA}
\affiliation{(present address) Department of Physics, University of California at
Berkeley, Berkeley, California 94720-7300, USA}

\author{Giacomo Saielli}
\affiliation{Istituto CNR per la Tecnologia delle Membrane, Sezione
di Padova, Via Marzolo, 1 - 35131 Padova, Italy}

\author{Alessandro Bagno}\email{alessandro.bagno@unipd.it}
\affiliation{Dipartimento di Scienze Chimiche, Universit\`a di Padova,
via Marzolo, 1 - 35131 Padova, Italy}

\author{Nhan Tran}
\affiliation{Department of Physics, Princeton University, Princeton,
New Jersey 08544, USA}

\author{Michael V. Romalis}
\affiliation{Department of Physics, Princeton University, Princeton,
New Jersey 08544, USA}

\date{\today}



\begin{abstract}
Scalar couplings between covalently bound nuclear spins are a ubiquitous feature in nuclear magnetic resonance (NMR) experiments, imparting valuable information to NMR spectra regarding molecular structure and conformation. Such couplings arise due to a second-order hyperfine interaction, and, in principle, the same mechanism should lead to scalar couplings between nuclear spins in unbound van der Waals complexes.  Here, we report the first observation of scalar couplings between nuclei in van der Waals molecules.  Our measurements are performed in a solution of hyperpolarized ${\rm ^{129}Xe}$ and pentane, using superconducting quantum interference devices to detect NMR in 10 mG fields, and are in good agreement with calculations based on density functional theory.  van der Waals forces play an important role in many physical phenomena, and hence the techniques presented here may provide a new method for probing such interactions.
\end{abstract}

\maketitle

\section*{Introduction}
Scalar coupling between nuclear spins, characterized by an interaction of the form
$H_J=hJ \mathbf{I_1}\cdot \mathbf{I_2}$, where $h$ is Planck's constant, $J$ is the
coupling constant, and $\mathbf{I_1}$ and $\mathbf{I_2}$ are the two spins involved, often referred to as
J-coupling, is a staple feature of
NMR spectroscopy.\cite{genref} J-coupling arises primarily
from a second-order Fermi-contact interaction, which
is generally thought to require covalent bonding between the two
coupled spins. In covalent molecules, this interaction gives rise to
observable splittings in NMR spectra whose pattern and separation
provides important information on molecular structure and conformation.
Such couplings vanish quickly upon increasing the
number of intervening bonds,
and couplings extending over more than 4-5 bonds are
generally very weak if at all observable. As a consequence, the
observation of a coupling constant is taken to be a signature of
covalent bonding, and is accordingly exploited to establish
molecular connectivities.

The two coupled nuclei are
therefore 
generally assumed to belong to
the same molecule. However, strictly there is no such requirement,
as long as the electronic structure of the system allows for the two
spins to be ``connected" by a (possibly low) electron
density.\cite{pnmrs} Very long-range couplings involving $^{19}$F
have been known for a long time and often denoted as occurring
``through space" (implying that the through-bond connectivity is too
long to allow for coupling to be transmitted).\cite{jff1,jff2} Hereafter,
we will use the term ``through space" even though the spin-spin
interactions involved are the same as in covalent
molecules.\cite{pnmrs} Such a situation holds also for even
relatively weak hydrogen bonds (HB), and through-HB couplings are
now routinely analyzed in structural biochemistry.\cite{jhb}
Similarly, through-space $J$($^{13}$C,$^{13}$C)
or $J$($^1$H,$^1$H) couplings have recently been reported for proteins whose conformation maintains a sufficiently small distance between coupled nuclei \cite{bryce} and for [2,2]paracyclophanes \cite{bour}.
Analysis of such couplings in molecules where the interacting spins are
connected by covalent bonds require careful consideration of
conformational effects \cite{Bagno2002} and coupling
pathways \cite{malkin} in order to assess their nature.

Scalar couplings in van der Waals molecules are expected to arise through the same mechanism as in the case of covalently bound spins. Pioneering computational
work on Xe$\cdots$Xe and Xe$\cdots$H \cite{harris} predicted very small couplings (in the range of $10^{-3}$ Hz). Further works using improved theoretical methods, both
\textit{ab initio} and based on density functional theory (DFT), have predicted couplings on the order of $10^{-1}$ Hz in
He$_2$, (CH$_4$)$_2$, a variety of organic
\cite{pecul_a,pecul_b,pecul_c,Bagno2001,Bagno2002} and xenon \cite{Bagno2003} complexes.
Similar calculations have enabled accurate prediction of
through-HB couplings.\cite{Bagno2000,jhb}

Experimental detection of couplings
in the range of $10^{-1}$ Hz is feasible with conventional high-field NMR techniques.  However, for unbound systems, chemical exchange and diffusion averages the effects of such coupling to very nearly zero in thermally polarized samples. Thus, despite the fundamental and practical implications of the existence of spin-spin couplings in unbound van der Waals molecules, there is no record of their experimental observation.

In this work, we report the first observation of scalar
J-couplings between unbound spins in van der Waals complexes in a solution of hyperpolarized liquid Xe and pentane. The approach is based on earlier observations
\cite{Heckman2003} that J-coupling in the presence of fast chemical
exchange can be observed if one spin species is hyperpolarized,
resulting in an average frequency shift of the other species. With
this technique, the strength of proton NMR signal is enhanced by a
factor of 10$^{6}$ by SPINOE with hyperpolarized
$^{129}$Xe.\cite{Navon1996,Fitzgerald1998} The measured value of J-coupling is in good agreement with density
functional calculations averaged over an ensemble of spins
simulating the bulk liquid phase. Van der Waals forces play an important role in many physical phenomena, and hence the ability to detect scalar couplings due to such forces may prove extremely valuable.  For example, since J-couplings in unbound systems
vanish quickly with distance, their detection and understanding
would provide a spatial restraint akin to the nuclear Overhauser effect.

\section*{Results and Discussion}
Consider a system of two different spins, 1 (pentane protons) and 2 (${\rm ^{129}Xe}$ nuclei).
%
%
%
The Zeeman contribution to the proton Hamiltonian in the presence of a magnetic field $\mathbf{B}=B_z\mathbf{\hat{z}}$ is $H_Z = -\hbar\gamma_1 \mathbf{I}_1\cdot\mathbf{B}_0$,
where $\gamma_1$ is the proton gyromagnetic ratio. In the presence of rapid chemical exchange, the normal J-coupling Hamiltonian, $h J\mathbf{I}_1\cdot\mathbf{I}_2$, averages to
$H_J = h \langle  J\rangle I_1\langle I_{2z}\rangle$,
where $\langle J\rangle$ represents the thermodynamically averaged coupling between $^{129}$Xe and $^1$H \cite{Heckman2003}, and we have assumed that the xenon spin polarization is nearly parallel to the magnetic field (note that $J$ is given in Hz here).  The Larmor precession frequency of protons (assumed to be positive) is thus $\nu_1 = (\gamma_1/2\pi)B_z+\Delta \nu_1$, where
\begin{equation}\label{dn}
\Delta\nu_1=-\langle J\rangle \langle I_{2z}\rangle
\end{equation}
and the minus sign originates from the difference in the signs of the Zeeman and J-coupling Hamiltonians.
$\langle J\rangle$ is given by \cite{Heckman2003}:
\begin{equation}
\langle J \rangle = n_2 \int J(r) \text{exp}(-V(r)/kT) d^3r
\label{integ1}
\end{equation}
where $n_2$ is the number density of spin $I_2$, $J(r)$ is a distance-dependent coupling constant, and $V(r)$ is an interatomic potential. Equation \ref{integ1} can be rearranged so as to highlight its relationship
with the bulk liquid structure, characterized by a radial
distribution function $g(r)$:
\begin{equation}
\langle J \rangle = n_2 \int_V 4 \pi r^2 J(r) g(r) dr. \label{integ}
\end{equation}

The frequency shift $\Delta\nu_1$ is analogous to that observed in optical pumping
experiments; in vapor mixtures of alkali metals and noble gases, spin polarization of the unpaired electron of the metal produces a shift in the NMR resonance of the noble gas. The shift is due to the Fermi Contact interaction between the unpaired electron of the metal and the nucleus of the noble gas, mediated by the hyperfine coupling constant.~\cite{Schaefer1989}  For the case of two nuclear spins the frequency shift can be parameterized as
\begin{equation}\label{Eq:fshift}
    \Delta\nu_{\rm 1} =
    \frac{\gamma_1}{2\pi}\kappa\frac{8\pi}{3}M_{2z},
\end{equation}
where $M_{2z}=\langle I_{2z}\rangle \mu_2n_2/I_2$ is the magnetization of spin 2 and $\mu_2$ is its magnetic moment ($\mu_2=-3.88 \times 10^{-24}~{\rm erg/G}$ in the case of ${\rm ^{129}Xe}$\cite{CRC}). $\kappa$ is a
dimensionless parameter representing the suppression or enhancement of the classical dipole magnetic field due to Coulomb repulsion or attraction of the particles.
Recalling Eq. \eqref{dn} and setting $I_2=1/2$, we have
\begin{equation}\label{Eq:J}
    \langle J \rangle = -\frac{8}{3}\kappa\gamma_1\mu_2 n_2.
\end{equation}

\begin{figure}
  \includegraphics[width=3.4 in]{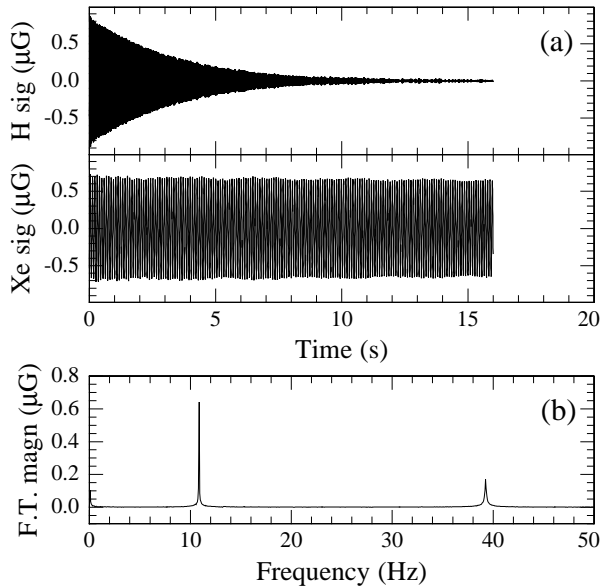}\\
  \caption{Typical signals obtained in our experiment after applying a $\pi/2$ pulse
  to the protons.  (a) Free induction decay signals for $^{129}$Xe and $^{1}$H after passing
  the raw signal through the appropriate bandpass filters. Proton $T_2$ was
  $3.2~{\rm s}$ for these data. (b) Magnitude Fourier transform of the raw signal. Owing to the very small applied field, proton chemical shifts in pentane are unresolved, yielding a single proton line at 39.2 Hz.}\label{Fig:Signals}
\end{figure}

\textbf{SQUID Measurements.} Our measurements are performed using hyperpolarized xenon (typical polarization is 2-3\%). Despite the use of hyperpolarized
$^{129}$Xe, frequency shifts of Eq. \ref{dn} are quite small, on the
order of tens of mHz, making detection in a high field spectrometer
challenging, due to the finite linewidth of resonances (either because of homogeneous or inhomogeneous broadening) and due to small magnetic field drifts.
We work in a low-field (10 mG), magnetically shielded
environment (see Supplementary Information, Fig. 1), where the high
absolute field stability facilitates observation of such small
frequency shifts. NMR signals are monitored
via high $T_c$ superconducting quantum interference devices (SQUIDs) to achieve high
sensitivity at low frequencies (magnetometric sensitivity was roughly ${\rm 0.3~ nG/\sqrt{Hz}}$).  Our measurements were performed using the following protocol: 1) The spherical sample cell was filled half full of pentane, which was then degassed via three freeze-thaw cycles under vacuum, and finally held in liquid state at about ${\rm -100^\circ C}$.  2) Hyperpolarized xenon gas was introduced into the sample cell where it condensed to liquid state to fill the remaining half of the cell. 3) Spin precession was excited by applying a
$\pi/2$ 
pulse to the protons via a rotating magnetic field at the proton resonance frequency.
Although this pulse was off 
resonance for the ${\rm ^{129}Xe}$ spins,
 it produced small excitation
, tipping them into the transverse plane by about
$1^\circ$.  
Proton $T_2$ was typically about 3.2 s,
 while ${\rm ^{129}Xe}$ $T_1$ was about 450 s,
so that many ($\approx 50$) proton transients could be collected for each batch of polarized xenon. Free induction decay signals for ${\rm ^{1}H}$ and ${\rm ^{129}Xe}$ nuclei are shown in Fig. \ref{Fig:Signals}(a), after passing the raw signal through bandpass filters centered about the Larmor precession frequencies at 39.2 and 10.8 Hz, respectively.
The magnitude Fourier transform of the raw signal is shown in Fig.~\ref{Fig:Signals}(b) where the $^{129}$Xe and proton resonances can be easily identified.

The signature of the desired interaction is a proton frequency shift that is linear in the xenon magnetization.  There are two sources of spurious frequency shifts that need to be accounted for. We first address dipolar fields associated with a deviation of the cell from perfectly spherical geometry (primarily due to the filling port).  These effects can be compensated for by acquiring data with the magnetic field oriented in three orthogonal directions.  We acquired three data sets with $\mathbf{B}_0$ oriented along each of the three directions $x$, $y$, and $z$.

\begin{figure}
  \includegraphics[width=3.4 in]{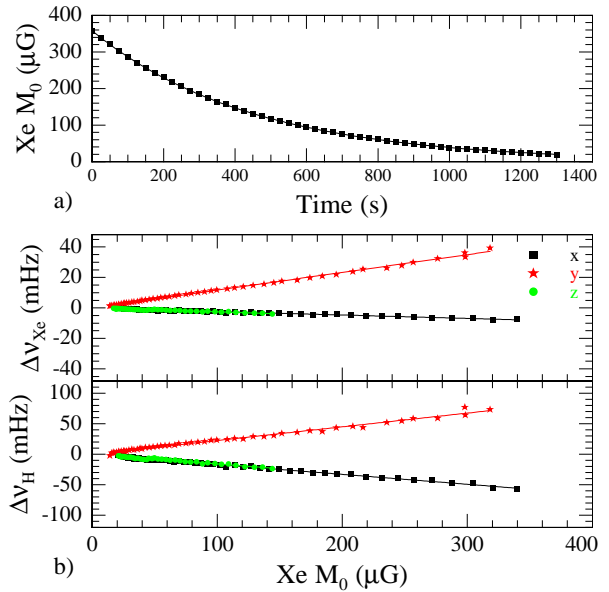}\\
  \caption{(a) Decay of longitudinal component of xenon magnetization, monitored by the
DC component of the SQUID signal in a typical data set. (b) Frequency shifts of ${\rm ^{129}Xe}$ and   ${\rm ^{1}H}$ NMR signals for the magnetic field oriented in three orthogonal directions. The straight lines overlying the data are linear fits.
  }\label{Fig:FreqShifts}
\end{figure}

Figure \ref{Fig:FreqShifts}(a) shows the decay of the longitudinal
component of the magnetization, determined from the DC value of the
SQUID signals.  Overlying these data is a decaying exponential with
time constant $T_1\approx 450~{\rm s}$. ${\rm ^{129}Xe}$ and proton resonance frequencies following each proton $\pi/2$ pulse were determined by passing the raw signal through the appropriate bandpass filter and fitting the result to a decaying sinusoid.  Figure \ref{Fig:FreqShifts}(b) shows the deviation of the ${\rm
^1 H}$ and ${\rm ^{129}Xe}$ NMR frequencies from their value in the
limit of low ${\rm ^{129}Xe}$ magnetization as a function of the ${\rm ^{129}Xe}$ magnetization.  Shown here are typical
data sets for $\mathbf{B}_0$ oriented in three different
directions.

The second source of spurious frequency shifts is
due to the magnetic fields associated with the pickup and feedback coils of the SQUID magnetometers.  The size of this effect can
be accurately determined based on the geometry of the detectors and
several auxiliary spin-precession measurements (see Methods and Supplementary Information). After
correcting for the effect of the SQUIDs, we verify that $^{129}$Xe
frequency shifts vanish (scalar spin-spin couplings between ${\rm ^{129}Xe}$ spins are not observable), and extract the part of the proton frequency
shift that is attributed to scalar coupling with $^{129}$Xe.

Frequency shifts are summarized in Table \ref{Table:fshifts}; the last row gives the
net frequency shift after subtracting the shifts due to the SQUID
magnetometers. The net ${\rm ^{129}Xe}$ frequency shift is
consistent with zero, and the average proton frequency shift per unit of magnetization along the three axes is
$\Delta\nu_1/M_{2z} = -52\pm 11~{\rm Hz/G}$, corresponding to $\kappa
= -0.0014 \pm 0.0003$. The cell was half-filled with $^{129}$Xe, corresponding to a density $n_2 = 7\times
10^{21}~{\rm cm^{-3}}$.\cite{Xenondata}  Employing Eq. \eqref{Eq:J}, we find $\langle J\rangle = -2.7\pm 0.6
~{\rm Hz}$.  We note that
accuracy and precision can be considerably improved by (a) an apparatus that overcomes some technical
limitations (cell distortions, variations in cell geometry due to
filling, and inaccurate measurement of the longitudinal
magnetization for certain configurations of the magnetic field); (b)
low-$T_c$ SQUIDs with better sensitivity ($\approx{\rm 0.01~nG/\sqrt{Hz}}$) than the
high-$T_c$ SQUIDs used here; (c) a higher xenon polarization than
the 2-3\% used in this work (polarization as high as 65\% has been
reported \cite{Hersman2006}).

The measured value of $\langle J \rangle$ cannot be directly compared with DFT-calculated coupling constants for a single Xe-pentane van der Waals complex, since in the bulk phase a variety of molecular arrangements exists; thus we have combined DFT calculations with a simulation of the liquid phase as follows.

\begin{table}
  \centering
\begin{tabular}{| c | c | c | c |}
  \hline
  $B_0$ direction & $\Delta\nu_1/M_{2z}$ & $\Delta\nu_2/M_{2z}$  & $\Delta\nu_{2,pure}/M_{2z}$
  \\
\hline
  x & $-160 \pm 10$ & $ -21 \pm 1  $ & $-13 $ \\
  y & $ 162 \pm 29$ & $ 106 \pm 12 $ & $96 $\\
  z & $-170 \pm 9 $ & $ -32 \pm 2  $ & $-31 $ \ \\
  x+y+z & $-168\pm 32$ & $53\pm 12$ & $53 $ \\
  \hline
  SQUID shift & -12 & 39 &  39 \\
  \hline
  net & $-156\pm 32 $ & $14\pm 12$ & $13$ \\
  \hline
\end{tabular}
  \caption{List of frequency shifts per unit of magnetization (Hz/G) for the magnetic field oriented in
  orthogonal directions. Errors are determined from the scatter of points over several trials, and are not given for pure xenon runs because we have collected only one data set for each direction.}\label{Table:fshifts}
\end{table}

\textbf{Computational studies.} Two separate sets of calculations were carried out: First, through-space couplings were calculated with relativistic DFT methods for a series of xenon-pentane
arrangements, thereby characterizing the most relevant portion of the $^{129}$Xe-$^1$H coupling surface, see Figure 2 of Supplementary Information. Second, we determined the radial distribution function $g(r)$ for the bulk liquid phase, which allows averaging of calculated coupling constants via Eq. \eqref{integ}.
Most calculated couplings are negative and lie between $-$8 and 0 Hz
(a few positive values are also found, as previously reported
\cite{Bagno2003}).

The magnitude of these couplings is consistent with the results of other computational works based on various \textit{ab initio} and DFT theoretical methods
\cite{pecul_a,pecul_b,pecul_c,Bagno2001,Bagno2002,Bagno2003}. However, our values are larger than those reported by Salsbury and Harris for Xe$\cdots$H \cite{harris} by 2-3 orders of magnitude. In this regard, we note
that this seminal paper \cite{harris} employed the
Thomas-Fermi-Dirac functional, which is based on the uniform electron gas model. This early functional is known to suffer from neglect of
electron correlation and inaccurate description of the electron
density found in molecular systems
\cite{cramer}.

The main trend of the calculated couplings features a roughly exponential
decay as the distance increases from ca. 3 to 4 \AA\
(See Supplementary Information, Figure 3).
Accordingly, the couplings from the second
solvation shell are mostly negligible. An orientational dependence
of the coupling constants is evident from the spread of points: some
relatively large couplings are predicted even when the Xe--H separation
is larger than 4.0 \AA. Similar anomalous distance dependences were
observed in Xe$\cdots$CH$_4$ \cite{Bagno2003} and in
 HF$\cdots$CH$_4$ \cite{pecul_a,pecul_b} complexes.
A detailed exploration of the
angular dependence would probably not provide important new
information, since we are only interested in the average properties
of the bulk liquid phase.
Calculated couplings in the range 3.0-4.3 \AA\ were fitted to an
exponential decay curve $J(r) = a e^{-br}$, with $a = -2.7867$ $\times$ 10$^4$
Hz, $b = 2.719$ \AA$^{-1}$.

\begin{figure}
\includegraphics[width=3.4 in]{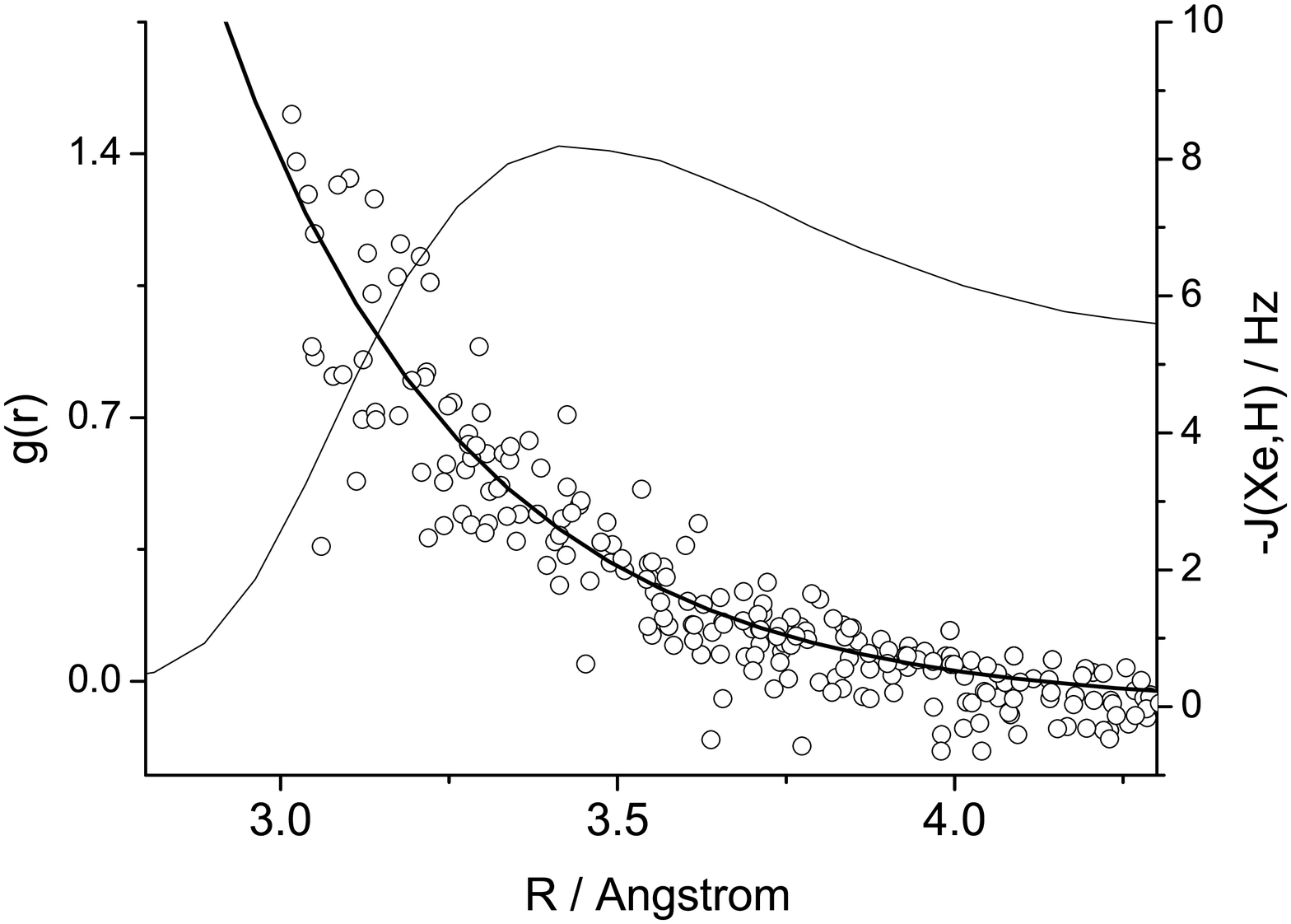}
\caption{(Thin line) Radial distribution function, $g(r)$, of the Xe--H
distance averaged over all protons of pentane, right axis. (Open circles)
calculated coupling constants with a negative sign for easier comparison with $g(r)$. (Thick line) fitting curve $-J(r) =
ae^{-br}$, with $a = -2.7867$ $\times$ 10$^4$ Hz, $b = 2.719$
\AA$^{-1}$, left axis.} \label{fig_int}
\end{figure}

In Figure~\ref{fig_int} we report the radial distribution function,
$g(r)$, of the distance between xenon and the hydrogen atoms of
pentane together with the calculated couplings and the fitting curve $-J(r)$.
There are three groups of chemically inequivalent protons in pentane, however since chemical shifts are unresolved in our experiment, we average the radial distribution function over all hydrogen atoms.
As we can see, the probability to find a Xe-H pair of atoms separated by less than 3 \AA \ is mostly negligible, thus there is no need to carefully characterize the $J(r)$ function for such short distances.
Integration of the
function of Eq. \ref{integ} with $n_2$ = 216/(31.76 \AA)$^3$ (see Simulation Details below and in Supplementary Information) gives
an average value $\langle J \rangle$ = $-$3.2 Hz, within
experimental error of SQUID measurements.

\begin{figure}
\includegraphics[width=3.4 in]{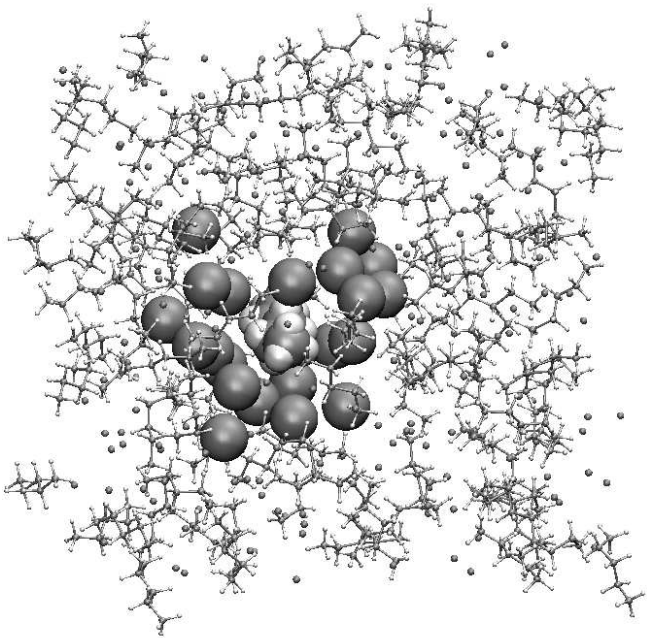}
\caption{Snapshot of the MD simulation box where we have highlighted a single pentane molecule and the closest xenon atoms more strongly contributing to the measured frequency shift of the pentane protons.} \label{fig_mdsnap}
\end{figure}

It is worthwhile to stress that the averaged coupling that we have determined is somewhat different from that encountered in high-field NMR spectra, where it is usually possible to measure the coupling constants between individual spin pairs. In contrast, the coupling we have determined between pentane protons and xenon is the sum of all coupling constants between protons and all xenon atoms surrounding pentane, since each coupling produces a shift which adds to the shifts of the others. This situation is illustrated by the MD snapshot in Figure \ref{fig_mdsnap}, where we have highlighted a single pentane molecule and the xenon atoms in the first solvation shell. Given the rapid decrease of $J(r)$ with distance, only the closest xenon shell provides a sizable contribution to $\langle J\rangle$. Thus, the value of $\langle J\rangle$ is close to that calculated at a Xe$\cdot\cdot\cdot$H distance of ca. 3.2 \AA.

For the same reasons, $\langle J \rangle$ depends on the density. Although the radial distribution function profile may also be density-dependent, the largest effect is simply a linear scaling of the coupling with the density of hyperpolarized xenon. Thus, in the limit of infinite dilution of pentane (cell filled with xenon), the frequency shift, and therefore $\langle J \rangle$, would be twice as large as in our case, where the mixture is only 50\% xenon.

\section*{Conclusions}
We have reported the observation of a scalar coupling between ${\rm
^{129}Xe}$ and ${\rm ^1H}$ in a solution of hyperpolarized xenon and
pentane. The measured value is in good agreement with calculations
based on density functional theory, averaged according to the bulk
liquid structure obtained by simulation. To the best of our
knowledge, this is the first observation of scalar coupling mediated
by van der Waals interactions in the presence of rapid chemical
exchange, i.e. for ``unbound" spins. Given the rapid decay of such
couplings with the xenon-probe distance, their observation provides
cogent information on the spatial proximity of unbound partners without complications
arising from relaxation effects, as is often the case with nuclear Overhauser effects.  Our measurements are facilitated by operating in very low magnetic fields, allowing us to easily vary the direction of the magnetic field $B_0$ over three orthogonal orientations to average away the effects of dipolar fields.  In principle magic angle spinning in high field may accomplish the same thing, although the necessity of vacuum tight mechanical connections to the xenon hyperpolarization system would make this challenging.  Implementation of pulse sequences such as WAHUHA (Waugh-Huber-Haberlein)\cite{Wahuha}, designed to mitigate the effects of dipolar fields may also enable high-field detection of such effects.

M.P.L., N.T. and M.V.R. designed and carried out SQUID experiments,
analyzed the data and wrote the paper. A.B. and G.S. designed and
carried out DFT calculations and MD simulations, analyzed the data
and wrote the paper.

\section*{Methods}
\textbf{Experimental SQUID measurements.} As typical for DC SQUID magnetometers \cite{Greenberg1998}, ours
consist of a SQUID loop with two Josephson junctions, a
superconducting pick-up coil and a normal feedback coil (inset (a),
Figure S1 in the Supplementary Information). A feedback circuit controls the current in the feedback coil
such that the SQUID is locked to
a 
fixed point of its $V(\Phi)$
curve. The current in the feedback coil serves as a measure of the
applied magnetic field, but also generates small magnetic fields
which can affect the precession frequency of the nuclear spins. We
consider separately the AC and DC magnetic fields generated by the
SQUID coils. The AC field is generated in response to the
oscillating $z$ component of the magnetization (inset (b), Fig. S1 in the Supplementary Information), and is in-phase with $M_z$. It can be shown from
Bloch equations that the AC in-phase field will cause a shift of the
$^{129}$Xe NMR frequency by slowly rotating the longitudinal
component (parallel to $\mathbf{B}_0$) of the Xe magnetization into
the transverse plane. However, it does not cause an appreciable
shift of the proton NMR frequency because the proton spins are
tipped by approximately $\pi /2$ relative to $\mathbf{B}_0$. In
contrast, the DC field will affect both spin precession frequencies.
It is important to take into account the fact that the SQUID
feedback circuit is reset after every NMR pulse, which results in a
much smaller DC field. We directly measured the AC and DC fields of
the SQUIDs at the location of the spin sample by carefully measuring
the dependence of the
 $^{129}$Xe 
NMR frequency on the applied uniform
magnetic field.

\textbf{DFT calculations and molecular dynamics simulation.}
Spin-spin coupling involving heavy-atom nuclei such as xenon are
affected by major relativistic effects,\cite{buhl} which have been
dealt with by the implementation of the Zero-Order
Regular Approximation (ZORA) method in the framework of DFT using the software package ADF.\cite{adf} Thus, we ran a series of
calculations of the through-space $J$($^{129}$Xe,$^1$H) coupling
constants between the protons in a pentane molecule and a xenon atom, at the scalar
ZORA BP86/TZ2P level, considering only the Fermi contact,
diamagnetic spin-orbit, and paramagnetic spin-orbit terms.\cite{bp,cpl} The
position of the xenon atom was varied over a cubic grid of points,
with the origin at the center of mass of the pentane molecule for a total of 1428 calculated coupling
constants (see Supplementary Information, Fig. 3). The radial distribution function in the bulk liquid phase
was obtained by molecular dynamics simulations, for which we used an
all-atom force field (OPLS-AA) for pentane~\cite{cp} and a
Lennard-Jones model for xenon~\cite{xe} (see the Supplementary
Information for details).

\section*{Acknowledgements}
Financial support from the University of Padova (PRAT CPDA045589) is gratefully acknowledged.  M.P.L. appreciates useful discussions with Stephan Appelt, Dmitry Budker, and Alex Pines.

\end{document}